\documentclass[fleqn,10pt]{wlscirep}
\usepackage[utf8]{inputenc}
\usepackage[T1]{fontenc}
\usepackage{slashed}
\usepackage{amsmath}
\usepackage{hyperref}

\title{Force, metric, or mass: Disambiguating causes of uniform gravity}

\author[1,*]{Yuan Shi}
\affil[1]{Lawrence Livermore National Laboratory, Livermore, California 94551, USA}

\affil[*]{shi9@llnl.gov}

\begin{abstract}
In addition to nonzero forces and nontrivial metrics, here I show that a nonconstant Higgs expectation value, which endows elementary particles with their masses, also leads to apparent universal particle accelerations and photon frequency shifts. When effects of the Higgs is attributed to spacetime curvatures, a spurious stress-energy tensor is required in Einstein's equation. On cosmological scales, the spurious density coincides with the observed dark energy density. 
On smaller scales, effects of the Standard Model Higgs gradients are unlikely observable except near compact astrophysical bodies. To estimate the experimental precision required to disambiguate causes of apparent accelerations, I compare distinct effects of the force, metric, and Higgs profiles that cause uniform acceleration of a test particle. 
When the acceleration is caused by a force, the motion of all particles are hyperbolic with the same acceleration. However, when the cause is a metric, only a one-parameter family of particles undergo hyperbolic motion. In comparison, when the cause is a Higgs gradient, the trajectory of all particles are hyperbolic, but the acceleration is larger when the particle's energy is higher.  
The discrepancies among the three causes are minuscule on laboratory scales, which makes experimental tests very challenging.
\end{abstract}
\begin{document}

\flushbottom
\maketitle
%
%
\thispagestyle{empty}

\section*{Introduction}
Einstein's theory of general relativity (GR) interprets gravity as spacetime curvatures \cite{Einstein1913entwurf,Einstein1915FieldEqu,MTWbook2017}. Constructed to recover Newtonian gravity in the weak-field limit, GR is initially supported by three additional evidences: gravitational redshift of photons \cite{Einstein1911einfluss,Pound60,Vessot80,Muller2010precision,Chou10}, perihelion procession of Mercury \cite{Newcomb1898,Freudlich1915,Einstein1915erklarung,Schwarzschild1916,Newhall1983}, and gravitational lensing of light \cite{Soldner1804,Einstein16,Dyson1920,VONKLUBER1960,Reasenberg79,Shapiro04}. After almost a century, other predictions of GR have finally been confirmed, owing to recent detection of gravitational waves \cite{Abbott16,GW170817} 
and an image of a black hole \cite{EHTC19}. The classical GR would be complete if it were not due to outstanding difficulties on large scales \cite{Weinberg89,Bertone2005particle,Frieman08}, which have motivated alternative theories \cite{Brans61,Khoury04,Will2018theory}. Most notably, the standard
cosmological model requires $\sim68\%$ dark energy and $\sim27\%$ dark matter \cite{Tegmark04,Komatsu2011seven,Ade2016planck}, which have so far evaded direct detection efforts \cite{PDG19}.

While the geometry that enters the Einstein's equation is classical, matter that curves the spacetime is quantum. At low energy, quantum matter gravitates mostly because of its mass, and in the Standard Model of particle physics, elementary particles acquire their masses through the Higgs mechanism \cite{Higgs64}. Although nucleons, which dominate the mass of ordinary matter, are composite particles whose masses are predominantly binding energy \cite{Durr2008ab,Borsanyi2015ab,Aoki2017,Yang18}, dimensionful scales are ultimately set by elementary particles. For example, the scale can be set by the pion decay constant, which is related through electroweak processes to mass of the W boson, which is an elementary particle. Therefore, the mass of hadronic matter is also proportional to the Higgs vacuum expectation value (VEV), even though the proportionality constant is more subtle.

In this paper, I relax the commonly made assumption that the Higgs VEV is a constant. A nonconstant Higgs expectation value introduces universal mass gradients to all particles and therefore leads to additional effects. To see why the Higgs VEV may deviate from a constant, consider Yukawa coupling between the Higgs $\phi$ and fermions $\psi$. Relevant terms in the Lagrangian ($\hbar=c=1$) are 
$\mathcal{L}=u^2\phi^2/4-\lambda\phi^4/4!-f\phi\bar{\psi}\psi$, where $u$ is the mass of $\phi$, $\lambda$ and $f$ are dimensionless coupling constants. 
When $\phi$ acquires a VEV of $v=u\sqrt{3/\lambda}$ at the minimum of its self potential, the fermion gains a mass term $m=f\langle\phi\rangle$. 
Conversely, when the number density $\langle\bar{\psi}\psi\rangle$ is nonzero, the source term distorts the expectation value $\langle\phi\rangle$.
In the perturbative regime, the minima are shifted to $\langle\phi\rangle\simeq \pm v-f\langle\bar{\psi}\psi\rangle/u^2$, which is significant only when the mass density $\rho=f\langle\phi\rangle\langle\bar{\psi}\psi\rangle\sim v^2u^2$. For the Standard Model Higgs \cite{PDG19}, since \mbox{$v\approx246$ GeV} and \mbox{$u\approx125$ GeV}, the relevant mass density \mbox{$\sim 10^{29}\; \text{kg/m}^{3}$} is extremely high, which seems to suggest that a constant VEV remains a good approximation.

However, already at much lower density, $\langle\phi\rangle$ can attain strongly distorted configurations due to a phase transition of the nonlinear classical field equation \cite{Shi2021nonperturbative}. For hadronic matter in three spatial dimensions, the critical density of the phase transition is \mbox{$\rho_c\sim 10^{14} R_{\text{m}}^{-1}\; \text{kg/m}^{3}$}, where $R_{\text{m}}$ is the scale length of the matter distribution in units of meters. Since $\rho_c$ decreases linearly with $R$, even a low density of matter can cause strong distortions of $\langle\phi\rangle$ when the scale length is large enough. 
For example, on the scale of neutron stars, \mbox{$R\sim 10$ km}, so \mbox{$\rho_c\sim 10^{10}\;\text{kg/m}^{3}$} is much lower than nuclear matter density, which suggests that neutron stars may strongly distort the Higgs VEV. 
As another example, in the Big-Bang cosmology, $R$ may be estimated by the Hubble length $c/H$. In a dust-dominated universe, $H=2/3t$, so the critical density of the phase transition is \mbox{$\rho_c\sim 10^{5} t_\text{s}^{-1}\;\text{kg/m}^{3}$}, where $t_\text{s}$ is the time since the Big Bang in units of seconds. In comparison, the critical density of the universe is \mbox{$3H^2/(8\pi G)\sim 10^9 t_\text{s}^{-2}\;\text{kg/m}^{3}$}, where $G$ is the gravitational constant. Comparing the two densities suggests that within the first few hours of the Big Bang, the Higgs VEV may be strongly distorted.

A spacetime-dependent Higgs VEV is analogous to a varying refractive index of an optical medium.
For example, in plasmas, photons become massive particles \cite{Anderson63,Shi2016effective}, whose dispersion relation $\omega^2=\omega_p^2+c^2k^2$ allows one to identify the plasma frequency $\omega_p^2=e^2n_0/\epsilon_0m_e$ as the photon mass.
Here, $e$ is the electron charge, $m_e$ is the electron mass, and $\epsilon_0$ is the permittivity of free space.
When the plasma density $n_0$ has spatial gradients, light rays bend due to refraction. 
Moreover, when the plasma density varies in time, photon frequencies change. 
Analogously, in the case of the Higgs expectation value, a spatial gradient changes the momentum and a temporal gradient changes the energy of massive particles.
If the gradients are present but one is unaware of them, then trajectories of particles would require some spurious forces or spacetime curvatures to explain.

In addition to altering trajectories of massive particles, a nonconstant Higgs expectation value also causes apparent photon frequency shifts.
In one scenario, we can measure the photon frequency using an atomic clock, whose rate is proportional to the Rydberg energy $\text{R}_\text{y}=m_ee^4/32\pi^2\epsilon_0^2\hbar^2$. 
Alternatively, we can measure the photon wavelength using a grating, whose size is proportional to the Bohr radius $a_\text{B}=4\pi\epsilon_0\hbar^2/m_ee^2$.  
Notice that when the electron mass is larger, clocks tick faster and rulers span shorter. Consequently, a photon appears to change color if its emitter and detector are located at two spacetime locations where the Higgs expectation values are different.

\begin{figure}[t]
	\centering
	\includegraphics[width=0.49\textwidth]{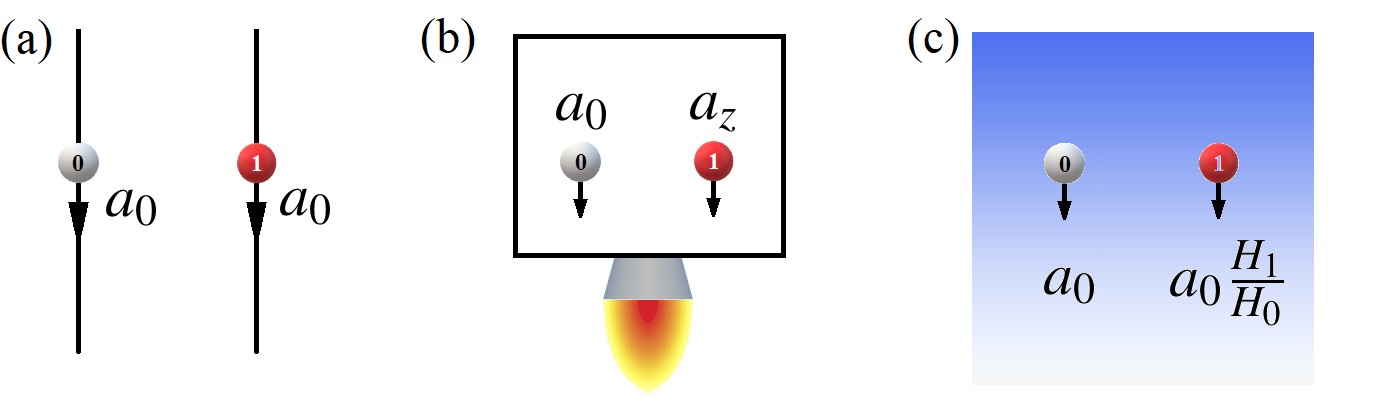}
	\caption{Apparent uniform acceleration due to force, metric, and mass may be disambiguated using more than one test particle.
	Even when a prototypical test particle (grey, ``0'') experiences a constant acceleration $a_0$ in its instantaneous rest frame, another test particle (red, ``1'') of the same composition but with different initial conditions may experience a different acceleration. 
	(a) For a constant force, $a_1=a_0$ is always the same constant. (b) For uniform frame acceleration, except for a one-parameter family, $a_1$ is not a constant but depends on $z$. (c) For a hyperbolic Higgs expectation value, although $a_1$ is always a constant, the acceleration is larger when the particle's energy is higher,
	}
	\label{fig:schematic}
\end{figure}

\section*{Results}
In what follows, I will consider the motion of test particles and show how a nonconstant Higgs expectation value introduces additional effects in gravitational theories. The force, metric, and Higgs profiles are regarded as prescribed, which do not react to the test particles. Since the mass $m$ of a particle is proportional to the Higgs expectation value, $m$ and $\langle\phi\rangle$ will be used interchangeably. $\hbar=c=1$ unless otherwise stated.

\subsection*{General theory}
When gradient scale lengths of the Higgs expectation value are much larger than the Compton wavelengths of test particles, geometric optics approximation applies and quantum wave packets are well described as classical point particles. Allowing its mass to vary, the trajectory of a classical particle is governed by the action (see Methods\ref{sec:Lagrangian} for a derivation)
\begin{equation}
	\label{eq:action}
	\mathcal{S}=-\int m \sqrt{dx^{\mu}g_{\mu\nu}dx^{\nu}}+eA_{\mu}dx^{\mu},
\end{equation}
which is manifestly invariant under change of coordinates. Here, the integration is along the world line of the particle, $g_{\mu\nu}$ is the metric tensor and $A_\mu$ is the electromagnetic gauge 1-form. While $g_{\mu\nu}$ and $m$ may have spacetime dependencies, the charge $e$ must be a constant due to gauge symmetry.

The classical trajectories are obtained by extremizing the action. To extremize the action, it is convenient to introduce an affine parameter $\xi$ along the particle's world line, such that $\mathcal{S}=\int d\xi L$. The Lagrangian is $L=-m(x^{'\mu}g_{\mu\nu}x^{'\nu})^{1/2}-eA_{\mu}x^{'\mu}$, where $x^{'\mu}=dx^\mu/d\xi$. 
The resultant Euler–Lagrange equations are simplified when written in terms of the proper time $\tau$, which is determined from $d\tau/d\xi=(x^{'\mu}g_{\mu\nu}x^{'\nu})^{1/2}$. 
The trajectories satisfy 
\begin{equation}
	\label{eq:EOM}
	\frac{d^2 x^\alpha}{d\tau^2} +(\Gamma^{\alpha}_{\mu\nu}+S^{\alpha}_{\mu\nu})\frac{dx^\mu}{d\tau}\frac{dx^\nu}{d\tau}
	=\frac{e}{m}g^{\alpha\mu}F_{\mu\nu}\frac{dx^\nu}{d\tau}+\frac{1}{\langle\phi\rangle}\frac{d\langle\phi\rangle}{d\tau}\frac{dx^\alpha}{d\tau}.
\end{equation}
In the above equation, $\Gamma^{\alpha}_{\mu\nu}=\frac{1}{2}g^{\alpha\sigma}(g_{\sigma\mu,\nu}+g_{\sigma\nu,\mu}-g_{\mu\nu,\sigma})$ is the usual Christoffel symbol, and $F_{\mu\nu}=A_{\nu,\mu}-A_{\mu,\nu}$ is the electromagnetic 2-form that gives rise to the Lorentz force. Here, $K_{,\mu}$ denotes $\partial_\mu K$ for a function $K(x)$ of the spacetime.

In the above equation of motion, there are two extra terms due to a nonconstant Higgs expectation value, one is analogous to a metric, whose effects are quadratic in 4-velocity, and the other is analogous to a force, whose effects are linear in 4-velocity. The $S^{\alpha}_{\mu\nu}$ tensor on the left-hand side (LHS) is given by 
\begin{equation}
	\label{eq:S}
	S^{\alpha}_{\mu\nu}=\frac{1}{\langle\phi\rangle}\Big(\delta^{\alpha}_\mu \langle\phi\rangle_{,\nu}+\delta^{\alpha}_\nu \langle\phi\rangle_{,\mu}- g_{\mu\nu}g^{\alpha\sigma}\langle\phi\rangle_{,\sigma}\Big).
\end{equation}
To see $S^{\alpha}_{\mu\nu}$ plays the same role as the Christoffel symbol under Weyl rescaling \cite{Weyl1918gravitation}, notice that we can pick a constant $m_0$ and redefine the metric to be $(m^2g/m_0^2)_{\mu\nu}$ in equation (\ref{eq:action}). Then, the Christoffel symbol $\Gamma^{\alpha}_{\mu\nu}[m^2g/m_0^2]=\Gamma^{\alpha}_{\mu\nu}[g]+S^{\alpha}_{\mu\nu}$. 
In other words, a spacetime dependent mass introduces a conformally flat component to the metric. In this sense, $\langle\phi\rangle$ plays a similar role as the scalar gravity \cite{Nordstrom1913,Ni75,Yang74}.
However, it is important to recognize that metric and mass play fundamentally different roles. To see why, notice that $\dot{x}^\mu g_{\mu\nu}\dot{x}^\nu=1$ is a constant of motion, where $\dot{x}^\mu:=dx^\mu/d\tau$. This identify can be verified directly using equation (\ref{eq:EOM}).
In contrast, $\dot{x}^\mu(m^2g/m_0^2)_{\mu\nu}\dot{x}^\mu=(m/m_0)^2$ is not a constant of motion unless $m$ is fixed.
This fact can be traced to the second term on the right-hand side (RHS) of equation (\ref{eq:EOM}), via which a nonconstant $\langle\phi\rangle$ leads to an effective force on the particle. This force is analogous to what a photon experiences in a refractive medium.
A more explicit form of equation (\ref{eq:EOM}) is $\ddot{x}^\alpha +\dot{x}^\alpha\dot{\langle\phi\rangle}/\langle\phi\rangle + \Gamma^{\alpha}_{\mu\nu}\dot{x}^\mu\dot{x}^\nu = e
F^\alpha_{\phantom{\mu}\nu}\dot{x}^\nu/m + \langle\phi\rangle^{,\alpha}/\langle\phi\rangle$. 
Since only ratios like $\langle\phi\rangle_{,\mu}/\langle\phi\rangle$ enter, effects of $\langle\phi\rangle$ are universal for all particles, regardless of their masses.

A spacetime-dependent $\langle\phi\rangle$ can mimic effects of a curved metric. Quantitatively, the metric connection defined by $(m^2g/m_0^2)_{\mu\nu}$ gives rise to a Riemann curvature tensor $R^{\mu}_{\nu\alpha\beta}[m^2g/m_0^2]=R^{\mu}_{\nu\alpha\beta}[g]+C^{\mu}_{\nu\alpha\beta}+r^{\mu}_{\nu\alpha\beta}$, where $C^{\mu}_{\nu\alpha\beta}=\Gamma^{\mu}_{\alpha\sigma}S^{\sigma}_{\nu\beta}+S^{\mu}_{\alpha\sigma}\Gamma^{\sigma}_{\nu\beta}-(\alpha\leftrightarrow\beta)$ and  $r^{\mu}_{\nu\alpha\beta}=S^{\mu}_{\nu\beta,\alpha}+S^{\mu}_{\alpha\sigma}S^{\sigma}_{\nu\beta}-(\alpha\leftrightarrow\beta)$. What is of importance in GR is the Ricci curvature tensor $R_{\mu\nu}=R^{\sigma}_{\phantom{\mu}\mu\sigma\nu}$, to which the Higgs expectation value contributes
\begin{equation}
	\label{eq:Ricci}
	r_{\mu\nu}=\frac{4\langle\phi\rangle_{,\mu}\langle\phi\rangle_{,\nu}}{\langle\phi\rangle^2}-\frac{2\langle\phi\rangle_{,\mu\nu}}{\langle\phi\rangle}-\frac{1}{2\langle\phi\rangle^2}\partial_\alpha\Big(g_{\mu\nu}g^{\alpha\beta}\langle\phi\rangle^2_{,\beta}\Big).
\end{equation}
Notice that $r_{\mu\nu}$ can be nonzero even when $g_{\mu\nu}$ is trivial. In other words, this effective curvature has little to do with the intrinsic geometry of the spacetime, and it arises only because one tries to interpret the contributions of $\langle\phi\rangle$ using a geometric language.
In the special case where the metric is Minkowski, the Ricci scalar $R=R^\mu_{\phantom{\mu}\mu}$ is given by the simple expression $r=-\frac{6}{\langle\phi\rangle}\partial^2\langle\phi\rangle$, where $\partial^2=\partial_\mu\partial^\mu$ is the d'Alembert operator.

\subsection*{Hyperbolic motion}
To illustrate the general theory, let us consider the conceptually important, albeit not realistic, example of hyperbolic motion  (figure \ref{fig:schematic}).
In Minkowski metric, the trajectory of a particle is hyperbolic in the lab frame if the particle experiences a constant force in its instantaneous rest frame \cite{Hill47}. Here, let us investigate what metric and Higgs expectation value can also lead to hyperbolic motion. 
Suppose the apparent acceleration is along the $z$ axis, then by symmetry, there exists a coordinate system, which I will refer to as the lab frame, in which the Lagrangian takes the form $-L=m\sqrt{U^2t^{'2}-V^2z^{'2}-x^{'2}-y^{'2}}+eA_tt^{'}$. Here, $m$, $U$, $V$, and $A_t$ are functions of $z$ only. The equations in $t$, $x$, and $y$ directions are therefore trivial, which give three constants of motion
\begin{eqnarray}
	\label{eq:H_constant}
	H&=& U^2 m\dot{t} +eA_t,\\
	p_x&=& m\dot{x},\\
	p_y&=& m\dot{y},
\end{eqnarray}
where dots denote derivatives with respect to the proper time. 
Only the $z$ equation is nontrivial, and the equation can be integrated. Alternatively, using the constant of motion $\dot{x}^\mu g_{\mu\nu}\dot{x}^\nu=1$, we can directly obtain the lab-frame velocity
\begin{equation}
	\label{eq:EOMz}
	\Big(\frac{d z}{d t}\Big)^2=\frac{U^2}{V^2}\Big[1-\Big(\frac{U}{H-eA_t}\Big)^2(p_x^2+p_y^2+m^2)\Big].
\end{equation}
Notice that the RHS is a function of $z$ only, so the above equation can be further integrated to determine a locally unique trajectory for given initial conditions.

In what follows, I will focus on the special case where the motion is along the $z$ direction, and consider effects of force, metric, and mass separately. Then, $p_x=p_y=0$, and the remaining initial conditions for a test particle ``0'' are $z(t=0)=z_0$ and $\beta(t=0)=\beta_0$, where $\beta:=dz/dt$ is the lab-frame velocity. Notice that in a nontrivial metric, $\gamma:=dt/d\tau$ is related to $\beta$ by $\gamma=(U^2-V^2\beta^2)^{-1/2}$, which recovers the usual expression in special relativity when $U=V=1$. The remaining constant of motion is $H_0=m_0\gamma_0U_0^2+eA_{t0}$, where $m_0$ denotes $m(z_0)$ and so on.

\subsubsection*{Constant force}
First, consider the familiar case of a constant force by setting $U=V=1$ and $m=m_0$. Then, equation (\ref{eq:EOMz}) becomes $\beta^2=1-1/[\gamma_0+e(A_{t0}-A_t)/m_0]^2$. For the trajectory of particle ``0 '' to be hyperbolic in the lab frame, we need the electric potential to be linear in $z$, namely,
\begin{equation}
	\label{eq:potential}
	A_t=A_{t0}-E_0(z-z_0),
\end{equation}
which corresponds to a constant electric field $E_0$ along the $z$ direction. 
With a constant electric force, the motion of another particle ``1 '', whose initial conditions are $z(t=0)=z_1$ and $\beta(t=0)=\beta_1$, is also hyperbolic with the same acceleration
\begin{equation}
	\label{eq:hyperbolic}
	[a_E^0(z-z_1)+\gamma_1]^2-(a_E^0 t+\gamma_1\beta_1)^2=1.
\end{equation}
Here, $a_E^0=eE_0/m_0>0$ is the bare acceleration due to the electric field. Since the metric is Minkowski, it is convenient to introduce rapidity $w$ such that $\beta=\tanh w$, then $\gamma=\cosh w$, $\gamma\beta=\sinh w$, and it is clear that the initial conditions are satisfied. The example trajectories due to a constant force are shown in figure \ref{fig:individual} (dotted).

\subsubsection*{Uniform metric}
Second, let us set $A_t=0$ and $m=m_0$, and consider what metric corresponds to uniform gravity in GR, which is indistinguishable from a uniform frame acceleration according to the equivalence principle. 
The only nonzero Christoffel symbols are $\Gamma^{t}_{tz}=U'/U$, $\Gamma^{z}_{zz}=V'/V$, and $\Gamma^{z}_{tt}=UU'/V^2$, where primes denote derivatives with respect to $z$. Then, components of the Riemann curvature tensor are either zero or proportional to $R^{t}_{\phantom{t}zzt}=(V/U)(U'/V)'$. The nonzero elements of the Ricci tensor are therefore
\begin{equation}
	\label{eq:Ricci_tensor}
	R_{tt}=\frac{U}{V}\Big(\frac{U'}{V}\Big)',\quad  
	R_{zz}=-\frac{V}{U}\Big(\frac{U'}{V}\Big)',
\end{equation}
and the Ricci scalar is $R=(2/UV)(U'/V)'$. 
For uniform gravity, the spacetime is flat, which occurs if the metric satisfies condition (i) $ U' \propto V$.
As is well noted by Rohrlich \cite{Rohrlich63}, this condition does not uniquely determine the metric. 
The metric can nevertheless be fixed by demanding that (ii) the trajectory of particle ``0 '' is hyperbolic, and (iii) the metric asymptotes to the Newtonian limit $U\simeq 1-a_G^0 (z-z_0)$ and $V\simeq 1$ when $z\rightarrow z_0$. Here, $a_G^0>0$ is the uniform frame acceleration, and the sign convention is such that the gravitational potential is lower for larger $z$. The unique metric satisfying these conditions is given by (see Methods\ref{sec:metric} for a derivation)
\begin{equation}
	\label{eq:metric}
	U=\frac{\gamma_0}{\cosh\rho},\quad V=\frac{\gamma_0}{\tanh w} \frac{\sinh\rho}{\cosh^2\rho}.
\end{equation}
In the above expressions, $\rho$ is related to $z$ by $\rho=\sqrt{[a_G^0(z-z_0)^2+\gamma_0]^2-1}+w_0-\sinh w_0$, and $w$ is related to $z$ by $\cosh w=a_G^0 (z-z_0)+\gamma_0$, where $w_0$ is the initial rapidity of particle ``0 '' such that $\cosh w_0=\gamma_0$.

Since the metric is flat, it can be trivialized to the Minkowski metric by suitable coordinate transformations. The transformations are not unique, but a convenient choice exploits $V=-U'/a_G^0$ to rewrite the metric $g=U^2dt^2-V^2dz^2=(U\cosh\theta dt+\sinh\theta dU/a_G^0)^2-(U\sinh\theta dt+\cosh\theta dU/a_G^0)^2$, where $\theta$ is an arbitrary phase.
The choice $\theta=a_G^0 t+w_0$ corresponds to the free-fall frame, in which particle ``0 '' remains at the origin. Explicitly, the lab frame $(t,z)$ and the free-fall frame $(\tilde{t},\tilde{z})$ are related by
\begin{eqnarray}
	\label{eq:t'}
	a_G^0 \tilde{t}&=&U(z) \sinh(a_G^0 t +w_0)- \sinh w_0,\\
	\label{eq:z'}
	a_G^0 \tilde{z}&=&\cosh w_0-U(z) \cosh(a_G^0 t +w_0),
\end{eqnarray}
where the sign choices are such that $\tilde{t}$ increases with $t$, and the lab origin recesses in the free-fall frame. 
The coordinate transformation becomes a Lorentz boost when \mbox{$a_G^0\rightarrow 0$}.
On the other hand, when $a_G^0$ is nonzero, the transform does not cover the entire spacetime. The Rindler horizon, which corresponds to $t=+\infty$, is located in the free-fall frame at
\begin{equation}
	\label{eq:Rindler}
	a_G^0(\tilde{z}+ \tilde{t})=\cosh w_0-\sinh w_0.
\end{equation}
Due to its perhaps inconvenient coordinate choices, the lab observer will never see any particle crossing the Rindler horizon nor receive any photon beyond the horizon, even though the spacetime is trivial.

\begin{figure}[t]
	\centering
	\includegraphics[width=0.5\textwidth]{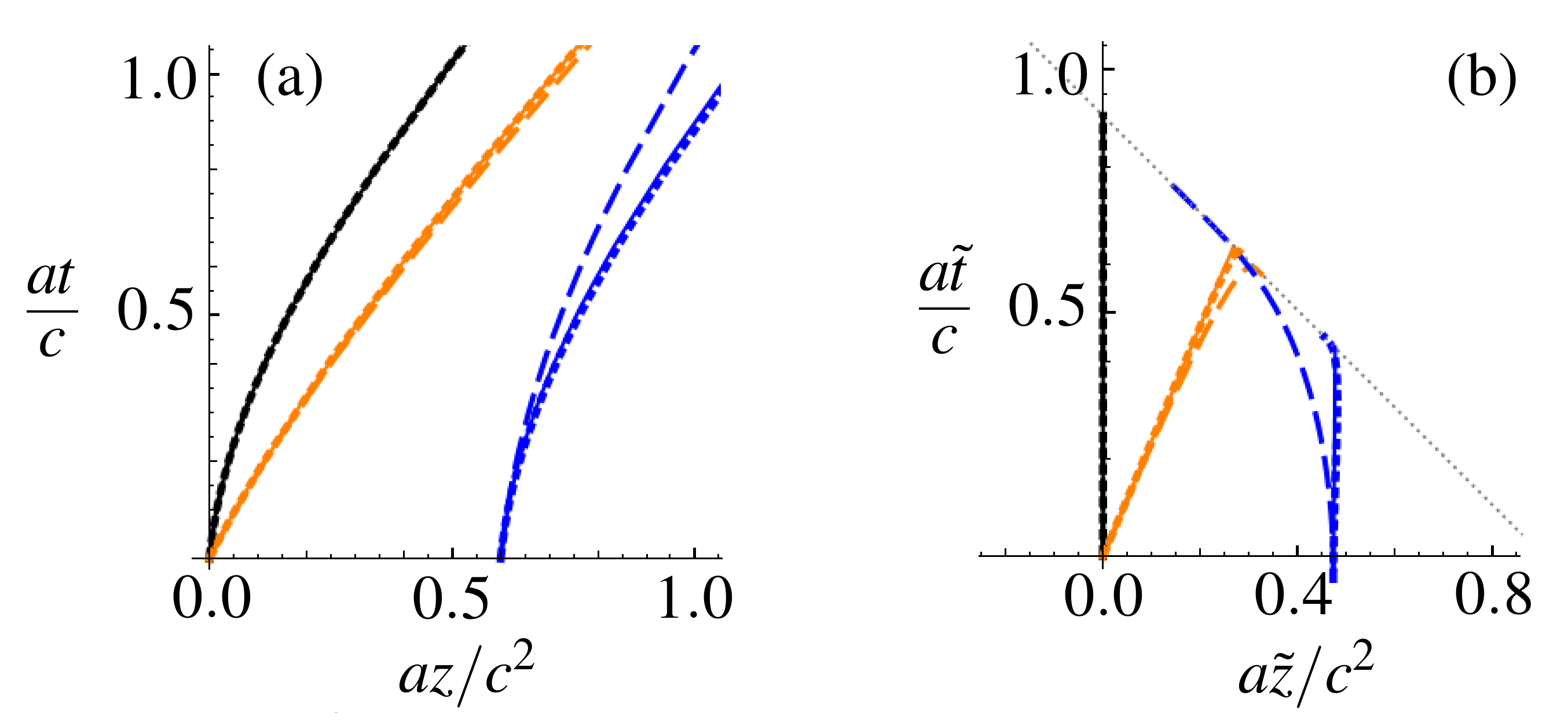}
	\caption{Example trajectories in lab frame (a) and free-fall frame (b) due to a constant force (dotted), a uniform frame acceleration (solid), or a hyperbolic Higgs expectation value (dashed). For the prototypical particle (black), these causes are indistinguishable. However, for another test particle with $z_1=0$ but $\beta_1=5\beta_0=0.5$ (orange), or $\beta_1=\beta_0$ but $z_1a/c^2=0.6$ (blue), the trajectories are different. The dashed gray line is the Rindler horizon. 
	}
	\label{fig:individual}
\end{figure}

Given the flat metric, geodesics are always straight lines in the free-fall frame, but they are not always hyperbolas in the lab frame. 
In the lab frame, equation (\ref{eq:EOMz}) can be written in terms of $\rho$ as $d\rho/dt=(a_G^0/\sinh\rho) (\cosh^2\rho-\epsilon_1)^{1/2}$, 
where $\epsilon_1=(\cosh^2\rho_1/\gamma_1\gamma_0)^2$ is determined by initial conditions of particle ``1 ''. When $\epsilon_1>0$, the trajectory is time-like
\begin{equation}
	\cosh\rho=\sqrt{\epsilon_1}\cosh(a_G^0 t+\theta_1),
\end{equation}
where $\cosh\theta_1=\cosh\rho_1/\sqrt{\epsilon_1}$. 
In the free-fall frame, the trajectory becomes the straight line
$a_G^0(\tilde{t} \sinh\Delta_1 -\tilde{z}\cosh\Delta_1)=(\cosh w_0-\cosh\rho_1)/\sqrt{\epsilon_1}$, where $\Delta_1=\theta_1-w_0$.
The geodesics are light-like when $\epsilon_1=0$, and space-like when $\epsilon_1<0$, whose equations are derived in Methods. 
Unlike the case of a constant force, uniform gravity in GR has a peculiar feature: The motion of a test particle is hyperbolic if and only if its initial conditions satisfy the constraint $\epsilon_1=1$, or equivalently $\beta_1^2=\tanh^2 w_1$. 
Apart from this one-parameter family, trajectories of other particles are not hyperbolic in the lab frame. On the other hand, any fixed point in the lab frame undergoes uniform acceleration in the free-fall frame, except that the acceleration $a_G^0/U_1$ depends on $z_1$. Example trajectories in both frames are shown in figure \ref{fig:individual} (solid).

\subsubsection*{Hyperbolic mass}
Finally, having discussed well-established causes of hyperbolic motion, let us consider effects of a nonconstant mass by setting $U=V=1$ and $eA_t=0$. Then, equation (\ref{eq:EOMz}) becomes
$(dz/dt)^2=1-(m/H_0)^2$, where $H_0=m_0\gamma_0$.
For particle ``0'' to undergo hyperbolic motion, the Higgs expectation value satisfies
\begin{equation}
	\label{eq:phi}
	\langle\phi\rangle=\frac{\gamma_0\langle\phi\rangle_0}{a_M^0 (z-z_0)+\gamma_0},
\end{equation}
where $a_M^0$ is the bare acceleration due to the mass gradient. When $a_M^0>0$ and $z>z_0$, the mass of any particle is depleted as it travels towards the positive $z$ direction. Since $\langle\phi\rangle$ has no time dependence, energy is conserved. Then, the momentum of the particle has to increase in order to compensate for the mass loss, leading to an apparent acceleration of the particle. 
At the turning point $z_T=z_0+(1-\gamma_0)/a_M^0<z_0$, particle ``0'' has zero velocity in the lab frame where all its energy is in the form of rest energy. 
At the singular point $z_S=z_0-\gamma_0/a_M^0<z_T$, $\langle\phi\rangle$ blows up. A similar singularity also occurs for $V$ in equation (\ref{eq:metric}) due to the demand for perpetual hyperbolic motion.

Given the form of the Higgs expectation value, trajectories of all particles are hyperbolic, but the acceleration depends on the particle's initial conditions. For particle ``1", its velocity satisfies $(dz/dt)^2=1-1/[a_1(z-z_S)]^2$, where $a_1=a_M^0 H_1/H_0$. Here, $H_1=m_1\gamma_1$ is the constant energy of particle ``1'', which is of the same composition as particle ``0'' but with a different initial position $z_1$ such that its initial mass is $m_1$. Integrating the velocity gives the hyperbolic trajectory 
\begin{equation}
	\label{eq:uniform_M}
	[a_1 (z-z_1)+\gamma_1]^2-(a_1 t+ \gamma_1\beta_1)^2=1.
\end{equation}
In contrast to the case of a uniform metric, where only a one-parameter family of particles undergo hyperbolic motion, in this case, the motion of all particles are hyperbolic. 
Moreover, comparing to the case of a uniform force, where all particles have the same acceleration, in this case, the acceleration is larger for particles of higher energy. This is intuitive because when the particle's energy is higher, it experiences a steeper $\langle\phi\rangle$ gradient due to Lorentz contraction, and therefore a larger acceleration in its instantaneous rest frame. 
Example trajectories due to the hyperbolic $\langle\phi\rangle$ are shown in figure \ref{fig:individual} (dashed).

\subsection*{Spurious stress-energy tensor}
When the acceleration caused by a nonconstant Higgs expectation value is falsely attributed to spacetime curvatures, spurious dark matter or dark energy may be required in order for Einstein's equation to hold. Consider the extreme case where the metric is trivial, then the Christoffel symbols due to the geometry are zero. In this case, the apparent Ricci curvature is entirely caused by gradients of $\langle\phi\rangle$, which is given by equation (\ref{eq:Ricci}).
If we enforce the Einstein's equation $r_{\mu\nu}-r g_{\mu\nu}/2-\Lambda g_{\mu\nu}=8\pi G T_{\mu\nu}$, where $\Lambda$ is the cosmological constant, then the spurious stress-energy tensor $T_{\mu\nu}$ has little to do with the presence of any matter or energy.
Although disturbance of the Higgs VEV may be caused by some local distributions of matter, gradients of $\langle\phi\rangle$ may extend well beyond the source region due to nonlinearities of the classical field equation \cite{Shi2021nonperturbative}. Moreover, even without any source, $\langle\phi\rangle$ is not guaranteed to take a constant value throughout the spacetime. In these scenarios, the spurious $T_{\mu\nu}$ may be nonzero even when the local matter and energy densities are zero.

To estimate the scale of the spurious stress-energy tensor, consider the example where the Higgs expectation value only has a gradient along the $z$ direction and the metric is Minkowski. Then, the only nonzero components of the Ricci tensor are
\begin{eqnarray}
	\label{eq:Ricci_txy}
	&&r_{tt}=-r_{xx}=-r_{yy}=\frac{\langle\phi\rangle''}{\langle\phi\rangle} + \Big(\frac{\langle\phi\rangle'}{\langle\phi\rangle}\Big)^2,\\
	\label{eq:Ricci_z}
	&&r_{zz}=3\Big[\Big(\frac{\langle\phi\rangle'}{\langle\phi\rangle}\Big)^2 -\frac{\langle\phi\rangle''}{\langle\phi\rangle}\Big],
\end{eqnarray}
and the Ricci scalar is $r=6\langle\phi\rangle''/\langle\phi\rangle$. If we enforce the Einstein's equation, then the stress-energy tensor corresponds to an anisotropic ideal fluid, whose nonzero components are $T_{tt}=-T_{xx}=-T_{yy}=[(\langle\phi\rangle'/\langle\phi\rangle)^2-2\langle\phi\rangle''/\langle\phi\rangle-\Lambda]/8\pi G$ and $T_{zz}=[3(\langle\phi\rangle'/\langle\phi\rangle)^2+\Lambda]/8\pi G$. 
When the cosmological constant is zero, $T_{zz}>0$ corresponds to an ordinary positive pressure. However, since $T_{tt}=-T_{xx}$ are of opposite signs, the fluid cannot be constituted of ordinary matter. In general, the spurious fluid may have features that resemble both dark matter and dark energy.

In the special case where the Higgs expectation value leads to hyperbolic motion of test particles, the spurious fluid becomes isotropic and contributes additively to the cosmological constant. 
One way to see this is by substituting the $\langle\phi\rangle$ profile given by equation (\ref{eq:phi}) into equations (\ref{eq:Ricci_txy}) and (\ref{eq:Ricci_z}). Since $\langle\phi\rangle'/\langle\phi\rangle=-1/(z-z_S)$ and $\langle\phi\rangle''/\langle\phi\rangle=2/(z-z_S)^2$, the Ricci tensor $r_{tt}=-r_{xx}=-r_{yy}=-r_{zz}=3/(z-z_S)^2$, and the spurious stress-energy tensor $-T_{tt}=T_{xx}=T_{yy}=T_{zz}=[\Lambda+3/(z-z_S)^2]/8\pi G$. We see the hyperbolic Higgs expectation value adds a positive contribution to the cosmological constant.
Conversely, we can solve for the $\langle\phi\rangle$ profile by demanding that the fluid is isotropic. Setting $r_{zz}=r_{xx}$ gives an equation for $\langle\phi\rangle$, which can be written as $2\langle\phi\rangle'^2=\langle\phi\rangle \langle\phi\rangle''$. From this equation, it is easy to see that $\langle\phi\rangle'/\langle\phi\rangle^2$ is a constant, so $\langle\phi\rangle\propto1/(z-z_S)$. Then, up to some initial conditions, equation (\ref{eq:phi}) gives a unique profile.

In more general cases, when the apparent acceleration is caused by gradients of the Higgs expectation value alone, the spurious dark energy or dark matter density may be estimated from the ideal fluid $T_{zz}\simeq 3/8\pi G(z-z_S)^2$ using
\begin{eqnarray}
	\label{eq:density_a}
	\rho_s&\simeq&\frac{3 a^2}{8\pi G c^2}\approx 2.0\times 10^{-8} a_{\text{SI}}^2\;\text{kg/m}^3,\\
	\label{eq:density_L}
	&\simeq&\frac{3 c^2}{8\pi G L^2}\approx 1.6\times 10^{26} L_{\text{m}}^{-2}\;\text{kg/m}^3,
\end{eqnarray}
where the speed of light $c$ is restored for clarity. Here, $a_{\text{SI}}$ is the apparent acceleration in units of \mbox{$1\;\text{m/s}^2$}, and $L_{\text{m}}$ is the gradient scale length of $\langle\phi\rangle$ in units of meters. To see whether $\rho_s$ is of any physical relevance, let us estimate its typical values on three different scales.
First, on cosmological scales, the expansion of the universe is associated with $a\simeq -c q H_0$, where $q\approx -0.5$ is the deceleration parameter and \mbox{$H_0\approx 70\;\text{km}\,\text{s}^{-1}\,\text{Mpc}^{-1}$} is the Hubble parameter. Then, using equation (\ref{eq:density_a}), the spurious density \mbox{$\rho_s\sim10^{-26}\;\text{kg/m}^{3}$} is on the scale of the observed dark energy density \cite{PDG19}. This is perhaps not surprising, because using equation (\ref{eq:density_L}) and taking $L\sim c/H_0$ to be the Hubble length, the spurious density $\rho_s=3H_0^2/8\pi G$ coincides with the critical density of the universe.
The coincidence alone does not imply that dark energy is due to a nonconstant Higgs expectation value. Nevertheless, the observed cosmological redshifts may contain contributions from a time-dependent $\langle\phi\rangle$, which may contribute to the Hubble tension \cite{Poli11}.
Second, on galactic scales, if we estimate the acceleration by that of the Sun around the Milky Way, then \mbox{$a\sim(250\;\text{km/s})^2/(8 \;\text{kpc})\approx2.5\times10^{-10}\;\text{m/s}^2$}, so equation (\ref{eq:density_a}) gives  \mbox{$\rho_s\sim10^{-27}\;\text{kg/m}^{3}$}. 
Alternatively, to estimate an upper bound, using equation (\ref{eq:density_L}) and taking \mbox{$L\sim10$ kpc} to be the luminous radius of the galaxy, then \mbox{$\rho_s\sim10^{-15}\;\text{kg/m}^{3}$}. 
In comparison, the inferred local dark matter density \cite{PDG19} is on the order of \mbox{$10^{-21}\;\text{kg/m}^{3}$}, which falls within the above estimates. 
Finally, on Earth scales, the surface acceleration is \mbox{$a \approx9.8\;\text{m/s}^2$}, then equation (\ref{eq:density_a}) gives \mbox{$\rho_s\sim10^{-6}\;\text{kg/m}^{3}$}, which is hardly noticeable compared to the density of air at sea level. On the other hand, when estimating $\rho_s$ using equation (\ref{eq:density_L}), even if we take \mbox{$L\sim6.4\times10^6$ m} to be the radius of the earth, the spurious density \mbox{$\rho_s\sim10^{12}\;\text{kg/m}^{3}$} is still enormous.

\subsection*{Residual acceleration}
From the examples of hyperbolic motion, we see that using only one test particle, it is not possible to distinguish whether its apparent acceleration is caused by a force, a metric, or a mass gradient. Nevertheless, by also observing other particles with different initial conditions, the underlying causes may be disambiguated.

To illustrate the idea, let us consider the case where a constant electric field, a uniform frame acceleration, and a hyperbolic Higgs gradient are simultaneously at play. In this general case, although finding an explicit solution to equation (\ref{eq:EOMz}) is difficult, it is straightforward to calculate Taylor series of the trajectories. 
To leading order in time, the trajectory of particle ``1'' is $z\simeq z_1+\beta_1 t+\frac{1}{2} a_1 t^2+O(t^3)$, where $a_1$ is its initial acceleration. To find the lab-frame acceleration, take time derivative on both sides of equation (\ref{eq:EOMz}). For one dimensional motion along the $z$ axis, $d^2z/dt^2=\beta d\beta/dz$ can be written as
\begin{equation}
	\label{eq:acceleration}
	\frac{d\beta}{dt}=\beta^2\Big(\frac{U'}{U}-\frac{V'}{V}\Big)-\Big(\frac{U^2}{V^2}-\beta^2\Big)\Big(\frac{U'}{U}+\frac{m'}{m}+\frac{eA'_t}{m\gamma U^2}\Big).
\end{equation}
In the special case where the potential is given by equation (\ref{eq:potential}), the metric is given by equation (\ref{eq:metric}), and the Higgs expectation value is given by equation (\ref{eq:phi}), the initial acceleration $a_1=a_E^1+a_G^1+a_M^1$ contains contributions from 
\begin{eqnarray}
	\label{eq:aE}
	a_E^1&=&\frac{a_E^0}{\gamma_1}\Big(1+\frac{a_M^0 h_1}{\gamma_0}\Big)\frac{\cosh\rho_1^2}{\gamma_0^2}\Big(\frac{\tanh w_1^2}{\tanh\rho_1^2}-\beta_1^2\Big),\\
	\label{eq:aG}
	a_G^1&=&a_G^0\Big[\frac{\tanh w_1}{\tanh\rho_1}+\beta_1^2(\text{csch}w_1^2 \,\text{sech} w_1-\coth w_1\coth\rho_1)\Big],\\
	\label{eq:aM}
	a_M^1&=&\frac{a_M^0}{\gamma_0+a_M^0 h_1}\Big(\frac{\tanh w_1^2}{\tanh\rho_1^2}-\beta_1^2\Big),
\end{eqnarray}
where $h_1=z_1-z_0$. 
Since the accelerations depend on both $h_1$ and $\beta_1$, particles with different initial conditions have different initial accelerations.
While the metric acceleration $a_G^1$ only depends on $a_G^0$, the mass acceleration $a_M^1$ depends on both $a_M^0$ and $a_G^0$, and the force acceleration $a_E^1$ depends on all three bare accelerations.

Unlike usual tests of the universality of free fall, where particles have different compositions but comparable initial conditions \cite{Williams2004progress,Schlamminger08,Poli11,Touboul2017microscope,Archibald2018universality}, here it is crucial that their initial conditions are distinct in order to disambiguate causes of the apparent acceleration. In other words, although the free fall is universal for all particles, how their identical acceleration depends on the initial heights and velocities reveals the underlying causes of the acceleration. 
To estimate the required experimental precision, consider typical scales of laboratory tests on Earth, where the apparent gravitational acceleration is \mbox{$a_\oplus\simeq 9.8\; \text{m/s}^2$}. Then, $a h_1/c^2\sim10^{-13} (h_1/1\,\text{km})$ is a small parameter, even when the height of the drop tower is on kilometer scale. Therefore, it is sufficient to expand equations (\ref{eq:aE})-(\ref{eq:aM}), treating $ah_1$ as a small parameter.

\begin{figure}[t]
	\centering
	\includegraphics[width=0.52\textwidth]{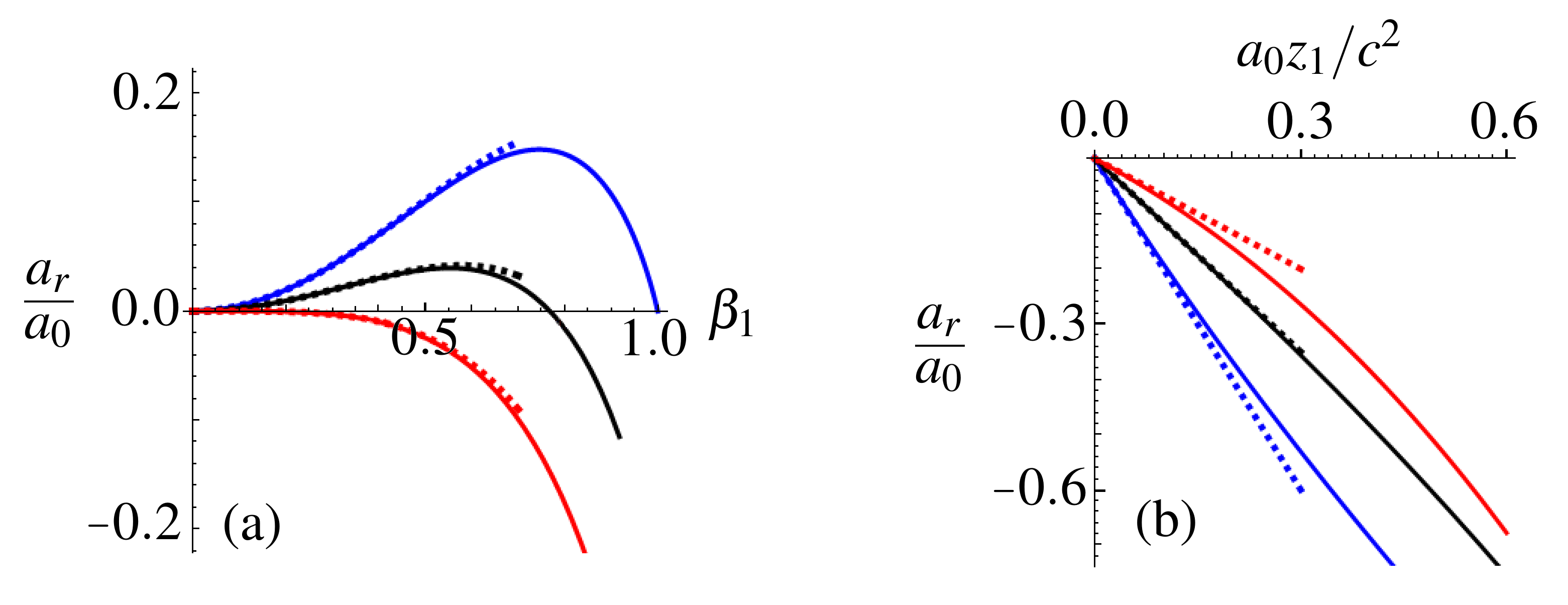}
	\caption{Exact (solid) and approximate (dashed) residual acceleration $a_r$ when an electric field is setup to cancel the acceleration at $z=z_0$ and $\beta_0=0$. 
	(a) Test particles are launched at the same height $z_1=z_0$ but with different initial velocities $\beta_1>0$. (b) Test particles are released from rest $\beta_1=0$ but from different heights $z_1>z_0$.
	When the uniform metric and the hyperbolic Higgs expectation value each contributes $+50\%$ (black) of the total acceleration $a_0$, the residual acceleration lies between the extreme cases of pure metric (red) and pure Higgs gradient (blue). 
	}
	\label{fig:residual}
\end{figure}

In the first scenario, we can launch particles with different velocities $\beta_1$ from the same height $h_1=0$. In this case, $U_1=V_1=1$, and $\gamma_1=(1-\beta_1)^{-1/2}$. Using $w_1=\rho_1=w_0$, we have $a_E^1=a_E^0/\gamma_1^3$, $a_G^1=a_G^0[1/\gamma_1^2-\beta_1^2/\gamma_0(\gamma_0+1)]$, and $a_M^1=a_M^0/\gamma_0\gamma_1^2$. 
In the special case $\beta_1=\beta_0$, the ratios $a_E^1/a_E^0=a_G^1/a_G^0=a_M^1/a_M^0=1/\gamma_0^3$ are identical. 
Suppose we set up a constant electric field to cancel the acceleration at $\beta_1=0$, which equals to $a_E^0+a_G^0+a_M^0/\gamma_0$, then for $\beta_1\ne0$, the charged particle ``1'' experiences a residual acceleration 
\begin{equation}
	\label{eq:az0}
	a_r = \Big(a_G^0+\frac{a_M^0}{\gamma_0}\Big)\frac{\gamma_1-1}{\gamma_1^3}-\frac{a_G^0\beta_1^2}{\gamma_0(1+\gamma_0)}.
\end{equation}
Consider the case $\gamma_0=1$. Suppose the uniform gravity is entirely due to GR (figure \ref{fig:residual}a, red), then $a_M^0=0$ and $a_r\simeq -3a\beta_1^4/8$ \mbox{$\approx-3.7\times 10^{-24} (a/a_\oplus) v_{300}^4\;\text{m/s}^2$}, where $v_{300}$ is the initial velocity in units of \mbox{$300 \;\text{m/s}$}. 
On the other hand, if the uniform gravity is entirely due to the Higgs gradient (figure \ref{fig:residual}a, blue), then $a_G^0=0$ and $a_r\simeq a\beta_1^2/2$ \mbox{$\approx4.9\times 10^{-12} (a/a_\oplus) v_{300}^2\;\text{m/s}^{2}$}. These two extreme cases are of opposite signs and scale differently with $\beta_1$.
From these estimates, we see that the residual acceleration is minuscule unless the velocity is close to the speed of light. However, when $\beta_1\sim 1$, the velocity change $\Delta\beta_1/\beta_1\sim ah_1/c^2\ll1$ is also challenging to measure.

In the second scenario, we can release test particles from rest $\beta_1=0$ at different heights $h_1$. In this case, $\gamma_1=\cosh\rho_1/\gamma_0$ is nontrivial in the presence of GR time dilation, but equations (\ref{eq:aE})-(\ref{eq:aM}) are much simplified. Using expansions of the metric (see Methods), suppose an electric field is setup to keep particle ``0'' at rest, then the residual acceleration for particle ``1'' is
\begin{equation}
	\label{eq:ab0}
	a_r\simeq -\frac{h_1}{c^2}\Big[2\Big(\frac{a_M^0}{\gamma_0}\Big)^2 
	+2a_G^0 \frac{a_M^0}{\gamma_0} 
	+(a_G^0)^2\Big(1-\frac{1}{\gamma_0(\gamma_0+1)}+\frac{\delta_{0,\beta_0}}{6}\Big)\Big].
\end{equation}
Notice that $a_r$ is not continuous when $\beta_0\rightarrow0$, at which $d\rho/dz$ becomes singular at $z=z_0$.
Consider the case $\gamma_0=1$. Then, in pure GR where $a_M^0=0$ (figure \ref{fig:residual}b, red), the residual acceleration $a_r\simeq-2a^2h_1/3c^2$ \mbox{$\approx-0.7\times 10^{-15} (a/a_\oplus)^2 h_{\text{m}}\;\text{m/s}^{2}$}, where $h_{\text{m}}$ is the initial height in units of meters. 
In contrast, if the apparent acceleration is entirely caused by the Higgs expectation value for which $a_G^0=0$ (figure \ref{fig:residual}b, blue), then $a_r\simeq -2a^2h_1/c^2$ \mbox{$\approx-2.1\times 10^{-15} (a/a_\oplus)^2 h_{\text{m}}\;\text{m/s}^{2}$}. 
The negative gradient means weaker gravity closer to the ground, which is opposite to the Earth's gradient due to its spherical shape. 
If one carefully setup an electric field gradient to cancel that of the Earth's gravity, then with respect to the suspended particle ``0'', another test particle separated by $h_1\sim 10$ m will displace on the order of $\Delta h_1\sim -0.1$ mm after \mbox{$t\sim 1$ day}.

\section*{Discussion}
In addition to trajectories of test particles, frequency shifts of photons may also be used to disambiguate causes of the apparent gravity. In GR, it is usually assumed that one can readout the intrinsic metric of the spacetime using ideal clocks and rulers. However, when the Higgs expectation value is not a constant, one has to face the possibility that clocks and rulers may operate differently across the spacetime. 
With a usual clock, such as an atomic clock whose rate is proportional to the electron mass via the Rydberg energy, and a usual ruler, such as a crystal lattice whose length is inversely proportional to the electron mass via the Bohr radius, it is difficult to distinguish effects of time dilation in GR versus mass shift caused by a local depletion of $\langle\phi\rangle$. 
The key of making the distinction is therefore using alternative clocks whose rate is $\not\propto \langle\phi\rangle$ or rulers whose length is $\not\propto \langle\phi\rangle^{-1}$.

One way of making a clock whose rate is not proportional to $\langle\phi\rangle$ is to use a plasma oscillator. The rate of a plasma clock is determined by the plasma frequency $\omega_p^2=e^2n_0/\epsilon_0m_e$, where plasma density $n_0$ introduces a separate dimensionful scale that can be measured using a Langmuir probe \cite{Hutchinson2002principles}. 
In a simplified treatment, a planar Langmuir probe, when biased sufficiently negatively, draws an ion saturation current $I=e n_0 A\exp(-\frac{1}{2})(T_e/m_i)^{1/2}$ from the plasma. Here, $A$ is the probe area and $m_i$ is the mass of singly charged plasma ions. The characteristic drift velocity is determined by the plasma sheath potential that is proportional to the electron temperature $T_e$, which introduces another dimensionful scale. From the probe and thermodynamics measurements, $\omega_p^2=[eI\exp(\frac{1}{2})/\epsilon_0m_eA](m_i/T_e)^{1/2}$ can be determined.
The current $I=V/R$ can be measured using the Ohm's law, where $V\propto\langle\phi\rangle$ can be determined by the Josephson voltage standard and $R$ can be determined absolutely using the quantized resistance \cite{Klitzing80}. Then, the remaining dependencies on the Higgs expectation value come from $m\propto\langle\phi\rangle$ and $A\propto\langle\phi\rangle^{-2}$. Plasma clocks thus constructed have a rate $\omega_p\propto\langle\phi\rangle^{5/4}$, which depend on $\langle\phi\rangle$ differently than atomic clocks.
The challenge is to build a plasma clock that is accurate enough to measure minuscule gravitational redshifts.

In this paper, the prescribed hyperbolic profile of the Higgs expectation value is an asymptotic solution to the classical field equation near domain walls \cite{Dashen74extended}. For the Lagrangian $\mathcal{L}=\frac{1}{2}(\partial\phi)^2+\frac{1}{4}u^2\phi^2-\frac{\lambda}{4!}\phi^4$, the classical field satisfies $\partial^2\phi=\frac{1}{2}u^2\phi-\frac{\lambda}{6}\phi^3$. Near domain walls, the static field configuration diverges, and a dominant balance of the classical field equation is $\partial^2\phi/\partial h^2\simeq \frac{\lambda}{6}\phi^3$, where $h$ is the distance from the domain wall. 
An asymptotic solution is $\phi\simeq\pm\sqrt{12/\lambda}h^{-1}$, which yields equation (\ref{eq:phi}) with the identifications $h=z-z_S$ and $a_M^0=\sqrt{\lambda/12}\gamma_0\langle\phi\rangle_0$. 
To give an estimate of $a_M^0$, suppose $\gamma_0=1$ and $\langle\phi\rangle_0=v$. Then, for the Standard Model Higgs with $u=v\sqrt{\lambda/3}$ \mbox{$\approx125$ GeV}, the typical acceleration $a=uc^3/2\hbar$ \mbox{$\approx2.9\times 10^{34}\;\text{m/s}^2$} is enormous.
We see that domain walls provide strongly confining potentials that immediately bounce back massive particles.

The Higgs expectation value profile realized in nature remains to be measured. In the Standard Model, the Higgs VEV is assumed to be a constant throughout the spacetime. This is probably a good approximation a few hours after the Big Bang and away from compact astrophysical bodies \cite{Shi2021nonperturbative}. 
However, beyond the Standard Model, if $\langle\phi\rangle$ is not the result of spontaneous symmetry breaking, then its profiles are more open.
As a speculation, notice that a locally depleted $\langle\phi\rangle$ may lead to distortions of galaxy spectralgraphs. For a given spectral line, a radial mass shift compounded with an azimuthal Doppler shift gives $\Delta f=f(r)[\beta_0+\beta_\theta (r)\sin\iota\cos\theta]$, where $\beta_0$ is the mean velocity of the galaxy whose inclination is $\iota$. Lopsidedness of this type have been observed, while not satisfactorily explained, for $>50\%$ galaxies
\cite{Shostak74,Cohen79,Baldwin80,Bosma81a,Richter94,Rubin99,Bournaud05,Jog2009,Van2011lopsidedness,Andersen2013}. 
A local depletion of $\langle\phi\rangle$ could also contribute to the confinement of gases and stars and mimic effects of dark matter halos.

In summary, a nonconstant Higgs expectation value introduces a universal mass gradient that mimics some but not all aspects of general relativity. By comparing force, metric, and mass, it is shown that although their effects may coincide, the causes of apparent accelerations may be disambiguated by observing trajectories of particles with different initial conditions, or measuring photon frequency shifts using clocks whose rates are not proportional to the Higgs expectation value. 
On Earth scale, the expected discrepancies are minuscule, so it is challenging to use laboratory experiments to determine whether a nonconstant Higgs expectation value plays any role. If the Higgs expectation value turns out to have spacetime dependencies, then it may lead to effects that resemble dark energy and dark matter, and contribute to Hubble tension and lopsidedness of galaxy spectrographs.

\section*{Methods}
\subsection*{Action of a classical particle with a varying mass\label{sec:Lagrangian}}
For simplicity, consider the Minkowski metric with diagonal elements $(1,-1,-1,-1)$. When the scale lengths of the Higgs expectation value $\langle\phi\rangle$ are much larger than the Compton wavelength of a test particle, the particle's wave function can be solved using the WKB approximation. The local dispersion relation is $k_\mu k^\mu=m^2$, where $k_\mu=p_\mu-eA_\mu$ is the kinetic 4-momentum, and $p_\mu=(H, \mathbf{p})$ is the canonical 4-momentum. This dispersion relation contains both particle and anti-particle states. To select out the particle state, which has positive energy, take positive square root and the Hamiltonian is 
\begin{equation}
	\label{eq:Hamiltonian}
	H=\sqrt{(\mathbf{p}-e\mathbf{A})^2+m^2} + e A_t,
\end{equation}
which is valid even when $m$ is not a constant. The anti-particle state corresponds to the negative square root, whose treatment is largely parallel to the particle state discussed below.

Given the Hamiltonian, the phase space dynamics is governed by the Hamilton's equation. The canonical coordinates satisfy $dx^i/dt=\partial H/\partial p^i=(p^i-eA^i)/k^0$. Here, $dx^i/dt$ is the lab-frame velocity, which is commonly denoted as $\beta^i$ in special relativity. Using this equation and the local dispersion relation, we have $k_0^2:=(H-eA_t)^2=k_0^2\beta^2+m^2$, so $k^0=m\gamma$ where $\gamma=(1-\beta^2)^{-1/2}$ is the usual relativistic gamma factor. 
Then, $p^i=\gamma m\beta^i+eA^i$ and $H=\gamma m+e A_t$, namely,
\begin{equation}
	\label{:eq:momentum}
	p^{\mu}=\gamma m \frac{dx^\mu}{dt}+ eA^\mu, 
\end{equation}
which recovers the familiar result in special relativity, except that now $m$ is no longer required to be a constant.
Using these expressions, the other Hamilton's equation can be simplified as $dp^i/dt=-\partial H/\partial x^i=e \beta^j\partial_i A^j-\partial_i m/\gamma -e\partial_iA_t$. This equation for the canonical momentum can be converted to an equation for the kinetic momentum 
\begin{equation}
	\label{eq:force}
	\frac{d(m\gamma \beta^i)}{dt}=e(\mathbf{E}+\boldsymbol{\beta}\times\mathbf{B})^i- \frac{1}{\gamma}\partial_i m.
\end{equation}
Notice that one the LHS, a spacetime dependent $m$ must be kept inside the time derivative, and its total time derivative is $d_t=\partial_t+\beta^i\partial_i$. On the RHS, the first terms is the usual Lorentz force, and the second term arises due to a spatial mass gradient. Using the relation between $\beta$ and $\gamma$, the equation governing the time component of the kinetic 4-momentum is 
\begin{equation}
	\label{eq:energy}
    \frac{d(m\gamma)}{dt}=e \mathbf{E}\cdot\boldsymbol{\beta} +\frac{1}{\gamma}\partial_t m.
\end{equation}
One the RHS, the first term is the usual Joule heating, and the second term shows that a time varying mass causes energy change to the particle. This is intuitive, because from the local dispersion relation $k_0^2=\mathbf{k}^2+m^2$ we see that for a particle traveling at a given kinetic momentum, an increase of the mass will lead to an increase of the kinetic energy.

To convert the Hamiltonian formulation, which separates out temporal and spatial components, to the Lagrangian formulation, take the inverse Legendre transform $L=p^i\beta^i-H$. Using the relation between canonical and kinetic momenta, the Lagrangian can be written as
\begin{equation}
	\label{eq:Lagrangian}
	L=-\frac{m}{\gamma} -eA_\mu\frac{dx^\mu}{dt},
\end{equation} 
which is valid even when $m$ is not a constant. 
The negative signs are significant. The first term on the RHS becomes $-m\gamma\simeq -m+m\beta^2/2$ when $\beta^2\ll1$, which recovers the correct sign for the kinetic energy in the nonrelativistic limit. Notice that a spacetime dependent $m$ plays the role of a potential energy. 
The second term can be written as $eA_\mu\beta^\mu=e(A_t-\beta^iA^i)$, which corresponds to the potential energy of a point charge in background electromagnetic fields.

While the Lagrangian is not manifestly invariant under general coordinate transformations, the action is. Integrating in time, the action $\mathcal{S}=\int dt L$ can be written as 
\begin{equation}
	\label{eq:action_Minkowski}
	\mathcal{S}=-\int m\sqrt{dt^2-d\mathbf{x}^2} + eA_\mu dx^\mu,
\end{equation}
where the integration is along the particle's world line. Notice that $dt^2-d\mathbf{x}^2=dx^\mu\eta_{\mu\nu}dx^\nu$ is due to the Minkowski metric. Generalizing the metric to an arbitrary $g_{\mu\nu}$, we thus obtain the action given by equation (\ref{eq:action}), which is manifestly invariant under general coordinate transformations and is also invariant under the gauge transformation $A_\mu\rightarrow  A_\mu+\partial_\mu\alpha$, assuming the boundary contributions vanish.

\subsection*{Metric leading to hyperbolic motion of a test particle\label{sec:metric}}
A unique metric $g=U(z)dt^2-V(z)dz^2$ exists under conditions (i) the metric is flat, (ii) the trajectory of a particle is hyperbolic, and (iii) the metric asymptotes to the Newtonian limit $U\simeq 1-a (z-z_0)$ and $V\simeq 1$ when $z\rightarrow z_0$. 
Here, $a$ stands for $a_G^0$ for brevity.
Using equation (\ref{eq:Ricci_tensor}), condition (i) is satisfied if and only if $V\propto U'$. The proportionality constant is determined from condition (iii), which gives
\begin{equation}
	\label{eq:metric_V}
	V=-\frac{U'}{a}.
\end{equation}
To find an equation for $U(z)$, use condition (ii). Notice that for a particle ``0 '' with initial conditions $z(t=0)=z_0$ and $\beta(t=0)=\beta_0=\tanh w_0$, its lab-frame trajectory is hyperbolic if and only if $\beta^2=1-1/[a(z-z_0)+\gamma_0]^2$. Here, $\gamma_0=\cosh w_0$ is the usual gamma factor because $U=V=1$ at $z=z_0$ according to condition (iii). 
On the other hand, the trajectory satisfies equation (\ref{eq:EOMz}), which is simplified to $\beta^2=(U/V)^2[1-(U/\gamma_0)^2]$ for one dimensional motion along $z$ with $m=m_0$ and $eA_t=0$.
Matching the two expressions for $\beta^2$ and using equation (\ref{eq:metric_V}) gives an equation for $U$:
\begin{equation}
	\label{eq:metric_U}
	a^2U^2\Big(1-\frac{U^2}{\gamma_0^2}\Big)=\Big(\frac{dU}{dz}\Big)^2\Big[1-\frac{1}{a^2(z-z_s)^2}\Big],
\end{equation}
where $z_s=z_0-\gamma_0/a$. To solve this ordinary differential equation, it is convenient to introduce $w$ such that $\cosh w=a(z-z_s)$. Then, $dU/dz=(a/\sinh w)(dU/dw)$ and the RHS of equation (\ref{eq:metric_U}) becomes $(a/\cosh w)^2(dU/dw)^2$.
To further simplify, it is convenient to introduce $\rho$ such that $U/\gamma_0=1/\cosh\rho$. Then, equation (\ref{eq:metric_U}) becomes
\begin{equation}
	\label{eq:rho_w}
	\Big(\frac{d\rho}{d w}\Big)^2=\cosh^2 w,
\end{equation}
which an be easily integrated. To determine the initial condition, notice that $U=1$ at $z=z_0$, so initially $w=\rho=w_0$. 
To see that the sign of $d\rho/dw$ is positive, comparing $dU/dz=-a$ when $z\rightarrow z_0$ with $dU/dz=-a(\gamma_0\sinh\rho/\cosh^2\rho\sinh w)(d\rho/dw)$. Then, the solution is $\rho=\sinh w-\sinh w_0+w_0$, which gives equation (\ref{eq:metric}). 
The change of variable $z\leftrightarrow\rho$ is given explicitly by $[a(z-z_0)+\cosh w_0]^2-(\rho+\sinh w_0-w_0)^2=1$, which is invertible when $a(z-z_0)>1-\cosh w_0$ and $\rho>w_0-\sinh w_0$.

\subsection*{Geodesics in lab and free-fall frames\label{sec:geodesics}}
For the metric given by equation (\ref{eq:metric}), geodesics in the $(t,z)$ plane satisfy equation (\ref{eq:EOMz}), with $m=m_0$ and $eA_t=p_x=p_y=0$. Taking positive square root, the equation in terms of $\rho$ is
\begin{equation}
	\label{eq:EOM_rho}
	\frac{d\rho}{dt}=\frac{a}{\sinh\rho}\sqrt{\cosh^2\rho-\epsilon_1},
\end{equation}
where $\epsilon_1=(\cosh^2\rho_1/\gamma_1\gamma_0)^2=\cosh^2\rho_1-\beta_1^2\sinh^2\rho_1 \coth^2w_1$ is determined by initial conditions of particle ``1'', and $a$ stands for $a_G^0$ for brevity.
Finding trajectories in the free-fall frame requires the inverse of equations (\ref{eq:t'}) and (\ref{eq:z'}), which can be written as
\begin{eqnarray}
	\label{eq:t}
	&&\tanh (at+w_0)=\frac{a\tilde{t}+\sinh w_0}{\cosh w_0-a\tilde{z}},\\
	\label{eq:z}
	&&\cosh\rho=\frac{\cosh w_0}{\sqrt{(\cosh w_0-a\tilde{z})^2- (a\tilde{t}+\sinh w_0)^2}},
\end{eqnarray}
For $t$ and $z$ to be real valued, $\tilde{t}$ and $\tilde{z}$ need to satisfy inequalities $\cosh^2w_0\ge(\cosh w_0-a\tilde{z})^2- (a\tilde{t}+\sinh w_0)^2>0$. The second inequality gives the Rindler horizon.

First, when $\epsilon_1>0$, the trajectory is a time-like geodesic. To show this, change variable $\chi=\cosh\rho/\sqrt{\epsilon_1}$. Then, equation (\ref{eq:EOM_rho}) becomes $d\chi/dt=a\sqrt{\chi^2-1}$, which can be easily integrated to give $\cosh\rho=\sqrt{\epsilon_1}\cosh(a t+\theta_1)$. The initial phase is $\cosh\theta_1=\cosh\rho_1/\sqrt{\epsilon_1}$. 
To see this is a time-like geodesic, transform into the free-fall frame. Using identities of hyperbolic functions, the trajectory can be written as $a (\tilde{t} \sinh\Delta_1 -\tilde{z}\cosh\Delta_1)=(\cosh w_0-\cosh\rho_1)/\sqrt{\epsilon_1}$,
where $\Delta_1=\theta_1-w_0$. This is a straight line in the $(\tilde{t}, \tilde{z})$ coordinate with a slope $\tilde{\beta}=d\tilde{z}/d\tilde{t}=\tanh\Delta_1$. Since $|\tilde{\beta}|<1$, the line is a time-like geodesic.

Second, when $\epsilon_1=0$, the trajectory is a light-like geodesic. In this case, equation (\ref{eq:EOM_rho}) becomes $d\rho/dt=a/\tanh\rho$, which can be easily integrated to give $\cosh\rho=\cosh\rho_1 \exp(at)$. 
To see this is a light-like geodesic, transform into the free-fall frame, wherein $a (\tilde{t} -\tilde{z})=\exp(w_0)(\cosh w_0/\cosh\rho_1-1)$. 
This is a straight line with a slope $d\tilde{z}/d\tilde{t}=1$, and therefore a light-like geodesic. The case $d\tilde{z}/d\tilde{t}=-1$ corresponds to the negative square root for equation (\ref{eq:EOM_rho}).

Finally, when $\epsilon_1<0$, the trajectory is a space-like geodesic. To show this, change variable $\chi=\cosh\rho/\sqrt{|\epsilon_1|}$, whereby equation (\ref{eq:EOM_rho}) becomes $d\chi/dt=a\sqrt{\chi^2+1}$. Integrating this equation gives $\cosh\rho=\sqrt{|\epsilon_1|}\sinh(a t+\theta_1)$,
where $\sinh\theta_1=\cosh\rho_1/\sqrt{|\epsilon_1|}$. 
To see this is a space-like geodesic, transform into the free-fall frame, where the trajectory becomes $a (\tilde{t} \cosh\Delta_1 -\tilde{z}\sinh\Delta_1)=(\cosh w_0-\cosh\rho_1)/\sqrt{|\epsilon_1|}$.
This is a straight line in the $(\tilde{t}, \tilde{z})$ coordinate with a slope $\tilde{\beta}=d\tilde{z}/d\tilde{t}=\coth\Delta_1$, where $\Delta_1=\theta_1-w_0$. Since $|\tilde{\beta}|>1$, the line is a space-like geodesic.

\subsection*{Asymptotics of the metric near the origin\label{sec:expansion}}
To approximate the residual acceleration, it is necessary to expand the metric near $z_1\sim z_0$. The small expansion parameter is $\epsilon=a_G^0(z_1-z_0)/c^2\ll1$. Since $\rho_1=[(\epsilon+\gamma_0)^2-1]^{1/2}-\sinh w_0 +w_0$, the behavior of $\rho_1$ is distinct for $\epsilon$ much larger or smaller than $\sinh w_0$.

When $\epsilon\ll\sinh w_0$, we have $\rho_1\simeq w_0+\epsilon/\tanh w_0$ and $w_1\simeq w_0+\epsilon/\sinh w_0$. Then, $\cosh\rho_1/\cosh w_0\simeq 1+\epsilon$ and $\sinh \rho_1/\sinh w_0\simeq 1+\epsilon/\tanh^2w_0$. Consequently, $\tanh \rho_1\simeq \tanh w_0+\epsilon/\sinh w_0\cosh w_0$. To find the residual acceleration when $\beta_1=0$, we also need $\tanh w_1\simeq \tanh w_0+\epsilon/\sinh w_0\cosh^2 w_0$, which gives $\tanh w_1/\tanh \rho_1\simeq 1-\epsilon/\gamma_0(1+\gamma_0)$.
Then, from equations (\ref{eq:aE})-(\ref{eq:aM}), the first order accelerations are $a_E^1/a_E^0\simeq 1+a_M^0h_1/\gamma_0+a_G^0h_1[1-2/\gamma_0(1+\gamma_0)]$, $a_G^1/a_G^0\simeq 1-a_G^0h_1/\gamma_0(1+\gamma_0)$, and $a_M^1/a_M^0\simeq 1/\gamma_0-[a_M^0+2a_G^0/(\gamma_0+1)]h_1/\gamma_0^2$.

On the other hand, when $\epsilon\gg\sinh w_0=0$, we have $\rho_1\simeq \sqrt{2\epsilon}(1+\epsilon/4)$ and $w_1\simeq \sqrt{2\epsilon}(1-\epsilon/12)$, where the square root indicates the singular behavior of $\rho(z)$ at $z=z_0$ and $w_0=0$. In this case, $\cosh\rho_1 \simeq 1+\epsilon$, $\sinh \rho_1\simeq \sqrt{2\epsilon}(1+7\epsilon/12)$, so $\tanh \rho_1\simeq \sqrt{2\epsilon}(1-5\epsilon/12)$. With $\tanh w_1\simeq \sqrt{2\epsilon}(1-3\epsilon/4)$, we have $\tanh w_1/\tanh \rho_1\simeq 1-\epsilon/3$.
Substituting these into equations (\ref{eq:aE})-(\ref{eq:aM}) and setting $\beta_1=0$, the first order accelerations are $a_E^1/a_E^0\simeq 1+(a_M^0+a_G^0/3)h_1$, $a_G^1/a_G^0\simeq 1-a_G^0h_1/3$, and $a_M^1/a_M^0\simeq 1-(a_M^0+2a_G^0/3)h_1$. 



\providecommand{\noopsort}[1]{}\providecommand{\singleletter}[1]{#1}%

\section*{Acknowledgements}
This work was performed under the auspices of the U.S. Department of Energy by Lawrence Livermore National Laboratory under Contract DE-AC52-07NA27344 and was supported by the Lawrence Fellowship through LLNL-LDRD Program under Project No. 19-ERD-038. 

\section*{Author contributions statement}
Y.S. is the sole author.

\section*{Additional information}
\subsection*{Code and data availability}
The computer programs and raw data supporting the findings of this paper are openly available at \url{https://github.com/seanYuanSHI/UniformGravity}.

\subsection*{Competing interests}
The author declares no competing interests. 


\end{document}